\documentstyle[nato,psfig]{crckapb}

\begin{opening}
\title{THE INFLUENCE OF NOVA NUCLEOSYNTHESIS \protect\\
       ON THE CHEMICAL EVOLUTION OF THE GALAXY}

\author{D. ROMANO}
\author{F. MATTEUCCI$^1$}
\institute{SISSA/ISAS\\
           Trieste, Italy}
\institute{$^1$ Dipartimento di Astronomia, Universit\`a di Trieste,\\ 
           Trieste, Italy}

\end{opening}

\runningtitle{THE INFLUENCE OF NOVA NUCLEOSYNTHESIS}

\begin{document}


\begin{abstract}
We adopt up-to-date yields of $^7$Li, $^{13}$C, $^{15}$N from classical novae 
and use a well tested model for the chemical evolution of the Milky Way in 
order to predict the temporal evolution of these elemental species in the 
solar neighborhood. In spite of major uncertainties due to our lack of 
knowledge of metallicity effects on the final products of explosive 
nucleosynthesis in nova outbursts, we find a satisfactory agreement between 
theoretical predictions and observations for $^7$Li and $^{13}$C. On the 
contrary, $^{15}$N turns out to be overproduced by about an order of magnitude.
\end{abstract}

\section{Introduction}

Classical novae are sporadically injecting material processed by explosive 
hydrogen-burning nucleosynthesis into the interstellar medium (ISM). Both 
theoretical and observational evidence suggests that novae, although they 
have probably processed less than $\sim$ 0.3\% of the Galactic matter, may 
be important producers of $^7$Li, $^{13}$C, $^{15}$N, $^{17}$O, $^{22}$Na, 
and $^{26}$Al (see Gehrz {\it et al.} 1998 for a review). The role of novae 
in the enrichment of the ISM has already been investigated by D'Antona \& 
Matteucci (1991) (hereafter DM) and Matteucci \& D'Antona (1991). In DM $^7$Li 
production from novae is computed according to the results of Starrfield {\it 
et al.} (1978) concerning the explosive nucleosynthesis in nova outbursts and 
those of D'Antona \& Mazzitelli (1982) concerning the nova secular evolution. 
The DM results suggest that novae should contribute to more than 50\% to the 
global $^7$Li production, that $^{13}$C and $^{15}$N are produced 
proportionally to $^7$Li in novae, and that no primary $^{13}$C should 
originate from low-intermediate mass stars in order to fit the observational 
constraints in the DM scenario.
\par
Here we reanalyse the influence of novae on the chemical evolution of the 
solar neighborhood, by using the results recently obtained by Jos\'e \& 
Hernanz (1998) (hereafter JH) from a grid of hydrodynamical nova models.

\section{The chemical evolution model}

The adopted chemical evolution model is that of Chiappini {\it et al.} (1997). 
In this model the Galaxy is formed by two distinct episodes of mass accretion: 
the first one builds the halo and thick-disk substructures; the second one, 
almost independent from the former and temporally delayed with respect to it, 
originates the thin-disk, mainly through accretion of matter of primordial 
chemical composition.
\par
The explosive nucleosynthesis from nova outbursts is included in the model 
under quite straightforward assumptions on nova progenitors. We assume that 
at any given time $t$ the rate of nova outbursts is a fraction $\alpha$ of 
the white dwarf (WD) formation rate at a previous time $t - \Delta t$ 
multiplyed by the number of outbursts per nova ($n$):
\begin{displaymath}
R_{outbursts}(t) = n\,\alpha\int_{0.8}^{8} \psi(t - \tau_{m} - \Delta t) 
\phi(m)\,dm.
\end{displaymath}
We assume that all stars with initial mass between 0.8 and 8 $M_\odot$ end 
their life as WDs. $\tau_{m}$ is the lifetime of a star of mass $m$ and 
$\Delta t$ = 1 Gyr is the delay time required for the WD to cool and the 
first nova outburst to occur (Romano {\it et al.} 1999). The parameter 
$\alpha$, constant in time, is fixed from the condition of reproducing the 
currently observed rate of nova outbursts in the Galaxy, namely 
$R_{outbursts}(t_{Gal})$ $\sim$ 30 $\pm$ 30\% yr$^{-1}$ (Della Valle 2000). 
The parameter $n$ = 10$^4$ is the total number of outbursts suffered by the 
average nova (Bath \& Shaviv 1978).

\subsection{Detailed nucleosynthesis prescriptions}

\subsubsection{$^7$Li synthesis}

The stellar sources of $^7$Li considered are: carbon stars, massive AGB stars, 
Type II supernovae (SNe) and novae. For stars with initial masses in the range 
2\,--\,5 $M_\odot$ (carbon stars) and 5\,--\,8 $M_\odot$ (massive AGB stars) 
we evaluate the masses ejected as newly produced $^7$Li following DM and 
Matteucci {\it et al.} (1995). In particular, in Model A and Model B we adopt 
the same prescriptions on $^7$Li as in the best model of Matteucci {\it et 
al.} (1995), whereas in Model C we set to zero the contribution from carbon 
stars and assume a lower $^7$Li production from massive AGB stars. 
\par
For stars with initial mass $M$ $>$ 10 $M_\odot$ we use either the whole 
yields of $^7$Be + $^7$Li tabulated by Woosley \& Weaver (1995) (Model A and 
Model B) or the same yields but reduced by a factor of 2 (Model C).
\par
Li production during thermonuclear runaways in novae is included in our model 
by averaging on 14 evolutionary sequences computed by JH, to which we refer 
for details about input physics, nucleosynthesis and related uncertainties. 
We assume that $\sim$ 30\% of nova systems accrete hydrogen rich matter onto 
the surface of ONeMg WDs, whereas the remaining accrete hydrogen rich matter 
onto the surface of CO WDs. One model (Model A) does not take into account 
novae as Li producers.
\par
A further $^7$Li source is identified in the Galactic cosmic ray (GCR) 
spallation on ISM nuclei. Therefore, we run a model (Model C + GCRs) in which 
we add the absolute $^7$Li yields from GCRs as given by Lemoine {\it et al.} 
(1998) (their Table 1, case with lower-bound spectrum and $x$ = 1).

\subsubsection{Carbon and nitrogen evolution}

\begin{table}[htb]
\begin{center}
\caption{ Models referring to CN isotope evolution (see text).}
\begin{tabular}{l c c}
\hline
Model&novae&low-intermediate masses\\
\hline
Model 1&no&RV (without hot bottom burning)\\
Model 2&no&RV (with hot bottom burning)\\
Model 3&no&HG (constant mass loss)\\
Model 4&no&HG (variable mass loss)\\
Model 5&yes (average)&RV (without hot bottom burning)\\
Model 6&yes (average)&HG (constant mass loss)\\
Model 7&yes (average)&HG (variable mass loss)\\
Model 8&yes (min $^{13}$C)&RV (without hot bottom burning)\\
Model 9&yes (min $^{13}$C)&HG (constant mass loss)\\
Model 10&yes (min $^{13}$C)&HG (variable mass loss)\\
\hline
\end{tabular}
\end{center}
\end{table}
In the mass range 0.8\,--\,8 $M_\odot$ we adopt either the old yields from 
Renzini \& Voli (1981) (hereafter RV) (their case with or without hot bottom 
burning; {\it i.e.} their case with or without primary $^{13}$C and $^{14}$N 
production) or the most recent ones from van den Hoek \& Groenewegen (1997) 
(hereafter HG) (their case with constant efficiency of mass loss along the 
AGB or their case with variable efficiency of mass loss along the AGB).
\par
The nucleosynthesis prescriptions in the mass range $M$ $>$ 10 $M_\odot$ are 
from Woosley and Weaver (1995) in all models.
\par
As far as CN production from novae is concerned, we run models without adding 
their contribution, models in which the average carbon and nitrogen production 
from 14 evolutionary sequences (JH) is included, and models in which only 
the evolutionary sequences corresponding to the minimum $^{13}$C production 
are considered. The different models are listed in Table 1.

\section{Results}

\subsection{A(Li) vs. [Fe/H]}

\begin{figure}[]
\hspace{2.4cm}
\psfig{figure=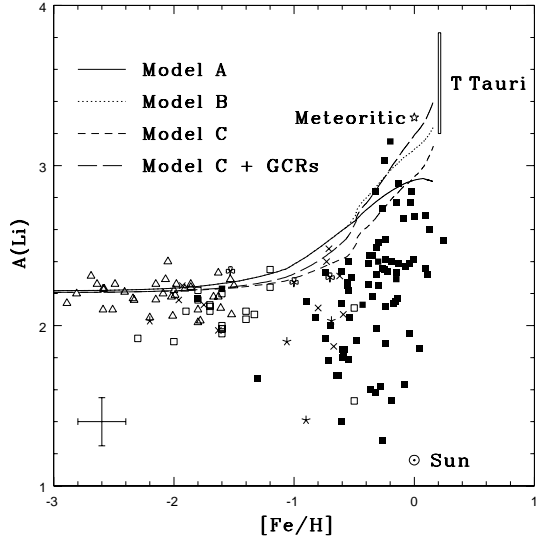,width=3in,height=3in}
\caption[]{\label{fig:fig 1} Theoretical predictions on A(Li) vs. [Fe/H] 
			     from four different models (prescriptions on
			     Li synthesis as in Sect.2.1.1) compared to 
			     observations.}
\end{figure}
The $^7$Li abundance is measured in stars of all metallicities and the 
observations indicate a plateau at low metallicities (PopII stars) followed 
by an increase in the $^7$Li abundance in the more metal-rich stars. However, 
in metal-rich stars $^7$Li can be strongly depleted (Fig.1) so the aim of 
chemical evolution models is to fit the upper envelope of the observed 
distribution, assuming that this represents the growth of $^7$Li in time 
as a consequence of stellar and perhaps GCR production. In this framework, 
the abundance of $^7$Li measured in PopII stars represents the primordial 
$^7$Li abundance, although a small contribution from GCRs already at these 
low metallicities cannot be excluded. A recent and careful analysis of extant 
data in the A(Li) vs. [Fe/H] diagram (Romano {\it et al.} 1999) has revealed a 
possible extension of the plateau toward metallicities higher than previously 
thought, [Fe/H] $\sim$ $-$\,0.4 dex. Therefore, astrophysical sources able to 
synthesize $^7$Li and restore it back into the ISM on long timescales should 
be preferred. Among these, novae are promising candidates: in fact, in our 
model the first WDs form after $\sim$ 600 million years from the beginning of 
Galaxy formation, then a billion year more has to elapse to allow them to cool 
at a level that ensures strong enough nova outbursts.
\par
In Fig.1 we show the theoretical A(Li) vs. [Fe/H] obtained from the four 
different models discussed in Sect.2.1.1. Models taking into account novae 
as Li factories (Models B, C, and C + GCRs) predict that a substantial amount 
of Li should be returned to the ISM at late times, thus reproducing the steep 
rise suggested by the data. In particular, Model C + GCRs guarantees a very 
good fit to the observational data but it also predicts a too high 
contribution from GCRs to the solar $^7$Li abundance ($\sim$ 45\%). In 
conclusion, we suggest that the best model should be intermediate between 
Model B and Model C + GCRs.

\subsection{The carbon isotope ratio}

\begin{figure}[]
\hspace{1.6cm}
\psfig{figure=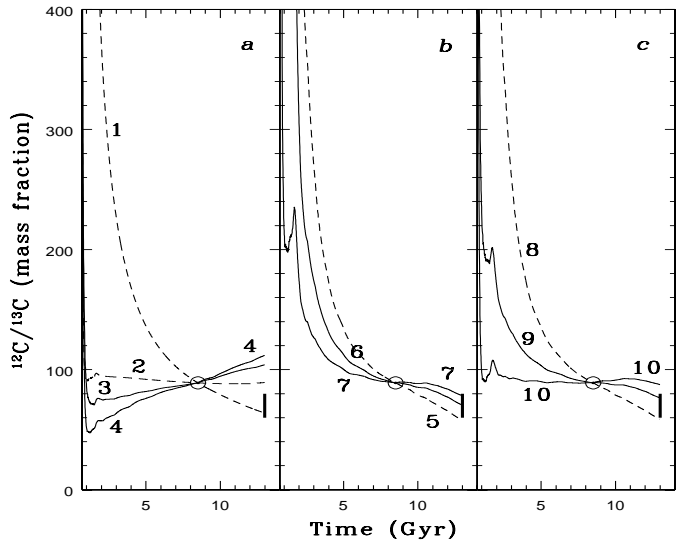,width=3.7in,height=3in}
\caption[]{\label{fig:fig 2} The temporal behaviour of the carbon isotope 
			     ratio as predicted by the model using different 
			     nucleosynthesis prescriptions (see Sect.2.1.2).
			     In all cases the model $^{12}$C/$^{13}$C ratio 
			     in the solar neighborhood 4.5 Gyr ago is 
		 	     normalised to its solar value.}
\end{figure}
\begin{figure}[]
\hspace{2.4cm}
\psfig{figure=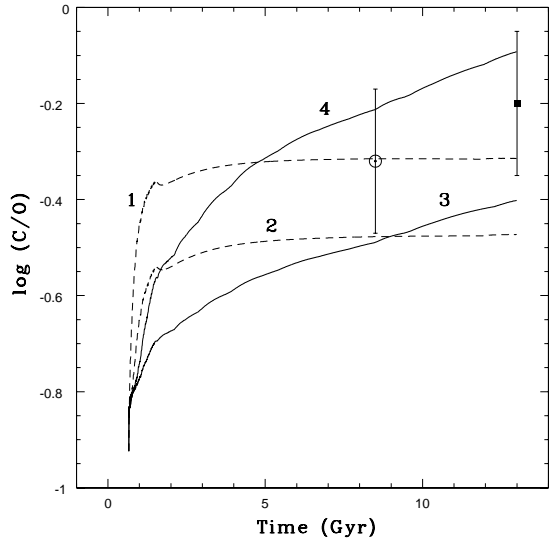,width=3in,height=3in}
\caption[]{\label{fig:fig 2} Theoretical log(C/O) vs. time from four different 
	                     models (see text for details).}
\end{figure}
The observations clearly show that the $^{12}$C/$^{13}$C ratio decreases from 
the time of the Solar System formation up to now: ($^{12}$C/$^{13}$C)$_\odot$ 
= 89 $\pm$ 2 (Cameron 1982), ($^{12}$C/$^{13}$C)$_{\mathrm{ISM}}$ = 40\,--\,80 
(Crane \& Hegyi 1988; Stahl {\it et al.} 1989). 
Values of ($^{12}$C/$^{13}$C)$_{\mathrm{ISM}}$ between 60 and 80 seem to be 
the most likely ones (Centuri\'on \& Vladilo 1991; Langer \& Penzias 1993).
\par
Among models excluding novae as CN isotope producers, Model 1 predicts too 
low $^{13}$C and $^{14}$N at the time of Sun's formation, whereas all the 
others are unable to properly fit the decreasing trend of the carbon isotope 
ratio in the last 4.5 Gyr (Fig.2, panel {\it a}). However, Model 4 predicts a 
log(C/O) vs. time in good agreement with observations (see Fig.3). This is due 
to the larger $^{12}$C production by low mass stars in the yields of HG 
compared to the old ones of RV. 
\par
Concerning $^{13}$C the overall situation is not yet clear, but an extra 
source of this element seems to be required to improve the agreement with 
observations.
\begin{table}[htb]
\begin{center}
\caption{ Elemental abundances (mass fraction) at the time of Sun's formation 
          as predicted by the model under different input nucleosynthesis, 
	  compared to the observations (Grevesse \& Noels 1993).}
\begin{tabular}{l c c c c}
\hline
{}&$^{12}$C$_\odot$&$^{13}$C$_\odot$&$^{14}$N$_\odot$&$^{15}$N$_\odot$\\
\hline
Model 1&$2.56E-3$&$8.89E-6$&$5.02E-4$&--\\
Model 2&$1.75E-3$&$4.29E-5$&$1.30E-3$&--\\
Model 3&$1.66E-3$&$1.39E-5$&$9.79E-4$&--\\
Model 4&$3.14E-3$&$2.56E-5$&$1.46E-3$&--\\
Model 5&$2.60E-3$&$9.10E-5$&$6.49E-4$&--\\
Model 6&$1.69E-3$&$9.75E-5$&$1.11E-3$&$2.44E-5$\\
Model 7&$3.17E-3$&$1.08E-4$&$1.59E-3$&$2.44E-5$\\
Model 8&$2.58E-3$&$3.11E-5$&$5.69E-4$&--\\
Model 9&$1.67E-3$&$3.74E-5$&$1.05E-3$&$2.49E-5$\\
Model 10&$3.15E-3$&$4.87E-5$&$1.53E-3$&$2.50E-5$\\
Observed&$3.62E-3$&$4.03E-5$&$1.07E-3$&$3.92E-6$\\
\hline
\end{tabular}
\end{center}
\end{table}
\par
$^{13}$C and $^{15}$N produced during a nova outburst are primary products 
since they form from C and O seeds which were synthesized during the previous 
evolution of the WD progenitor, starting from H and He. However, due to the 
long time taken for novae to expell their synthesized products, the abundances 
of $^{13}$C and $^{15}$N produced by novae behave like secondary products 
(Wilson \& Matteucci 1992, and refs. therein). As a consequence of this, when 
including this late contribution into the chemical evolution model, the 
$^{12}$C/$^{13}$C ratio in the solar neighborhood is expected to decrease from 
the time of the Solar System formation up to now. In Models 5, 6, and 7, 
where the nova contribution is added by averaging on the results from 14 
evolutionary sequences (JH), $^{13}$C turns out to be always overproduced (see 
Table 2). Therefore, we need some {\it ad hoc} assumptions: the average 
$^{13}$C production from a single nova system during the overall evolutionary 
history of the solar neighborhood should be not larger than the minimum one 
suggested by JH (Models 8, 9, 10). A model intermediate between Model 9 and 
Model 10 should guarantee the best fit to all the observational constraints 
considered here (see Table 2 and Fig.2, panel {\it c}). Unfortunately, the 
solar $^{15}$N abundance turns out to be overestimated by all the models.
\par
Therefore, we explain the decreasing $^{12}$C/$^{13}$C ratio in the local ISM 
as due to the primary nature of $^{12}$C coupled to the primary + secondary 
nature of $^{13}$C (see also Prantzos {\it et al.} 1996). A recent work by 
Palla {\it et al.} (1999) strongly supports the idea that the majority of the 
planetary nebula progenitors must have undergone an extra-mixing process, 
resulting in an enhanced $^{13}$C abundance in the surface layers and leading 
to a $^{12}$C/$^{13}$C ratio lower than predicted by standard stellar 
evolutionary models. This $^{13}$C source is equivalent to that from novae 
since in both cases this extra source is acting at late times. 
In summary: we cannot conclude that novae are necessary to explain the 
$^{12}$C/$^{13}$C ratio, but we can conclude that the inclusion of nova 
nucleosynthesis in the chemical evolution model for our galaxy improves 
the predictions for both $^7$Li and $^{13}$C evolution.


\begin{acknowledgements}
We thank the organizers for a very interesting meeting and C. Chiappini and 
N. Prantzos for many helpful comments. We acknowledge financial support from 
the Italian Ministry for University and for Scientific and Technological 
Research (MURST).
\end{acknowledgements}

\end{document}